# Electromagnetic response in spiral magnets and emergent inductance

Daichi Kurebayashi[1]*, Naoto Nagaosa[1,2]

[1] Center for Emergent Matter Science, RIKEN, Wako 351-0198, Japan
[2] Department of Applied Physics, University of Tokyo, 7-3-1 Hongo, Bunkyo-ku, Tokyo 113-8656

Corresponding author. Email: daichi.kurebayashi@riken.jp

**Abstract**
Emergent electromagnetism in magnets originates from the strong coupling between conduction electron spins and those of noncollinear ordered moments and the consequent Berry phase. This offers possibilities to develop new functions of quantum transport and optical responses. The emergent inductance in spiral magnets is an example recently proposed and experimentally demonstrated, used the emergent electric field induced by alternating currents. However, the microscopic theory of this phenomenon is missing, which should reveal the factors to determine the magnitude, sign, frequency dependence, and nonlinearity of the inductance L. Here we theoretically study electromagnetic responses of spiral magnets taking into account their collective modes. In sharp contrast to the collinear spin-density wave, the system remains metallic even in one-dimension, and the canonical conjugate relation of uniform magnetization and phason coordinate plays an essential role, determining the properties of L. This result opens a way to design the emergent inductance of desired properties.

**MAIN TEXT**

**Introduction**

Quantum transport phenomena in magnets include variety of effects such as magnetoresistance (1-3), planar Hall effect (4), spin-dependent tunneling (5), anomalous Hall effect (6-10), and spin Hall effect (11-16). Spin-orbit interaction is often relevant to these phenomena because the transport reflects the orbital motion of electrons while the magnetism comes from electron spins. In addition to the conventional ferromagnetism and antiferromagnetism, the recent focus is the noncollinear spin structures, which are induced by several mechanisms, e.g., frustrated exchange interaction (17), RKKY interaction (18-24), Fermi surface nesting (25,26), and Dzyaloshinskii-Moriya (DM) spin-orbit interaction (27-29). Note that there are situations where the system shows noncollinear spins without the spin-orbit interaction, which we study below. Noncollinear spin configurations are associated with the quantal Berry phase (30), which produces the emergent electromagnetic field (EEMF) (31-33). Emergent electric field $e$ is defined as

$$e_i = -\frac{1}{c}\frac{\partial a_i}{\partial t} = \frac{\hbar}{2e}(\boldsymbol{n} \cdot \partial_i \boldsymbol{n} \times \partial_t \boldsymbol{n}), \qquad (1)$$

while the emergent magnetic field $b$ by

$$b_i = (\boldsymbol{\nabla} \times \boldsymbol{a})_i = \frac{\hbar}{2e}\varepsilon_{ijk}(\boldsymbol{n} \cdot \partial_j \boldsymbol{n} \times \partial_k \boldsymbol{n}), \qquad (2)$$

where $\boldsymbol{n}$ is the direction of the spin. Note here that $b$ is associated with the noncoplanar spin structure such as the skyrmion, while $e$ is induced by the dynamics of spins. EEMF is the origin of many electromagnetic phenomena including the topological Hall effect by $b$ (34-37) and emergent electromagnetic induction by $e$ (38).

A helical or spiral spin structure with a single wavevector $\boldsymbol{Q}$ is a noncollinear but coplanar structure with $\boldsymbol{b} = \boldsymbol{0}$ in the ground state, while $e$ is induced by its dynamics. Recently, it has been proposed that the AC current-driven motion of the spiral leads to the inductance $L$ which is inversely proportional to the cross section $A$ of the sample in sharp contrast to the conventional inductor with $L$ being proportional to $A$ (39,40). The mechanism is based on the spin transfer torque; the angular momentum transfer between the conduction electrons and magnetic structure drives the motion of the latter, which produces the emergent electric field given by Eq.(1). In this picture, the quantum dynamics of the conduction electrons is not treated on a microscopic basis, and only their current density $j$ appears in the analysis.

Experimentally, the inductance $L$ of a short-period helimagnet $Gd_3Ru_4Al_{12}$ has been observed (41). The value of $L$ is around $\sim 100$ nH, which is comparable to the best commercial value while the size of the sample is $\sim 10^5$ smaller. Here there appeared several issues. One is the sign of the inductance $L$. Usually, the magnetic energy induced by the current is given by $\frac{1}{2}LI^2$ ($I$: current), and the negative $L$ means that the system is unstable. Therefore, it is urgent to understand what determines the sign of $L$ and its relation to the stability of the system. The second important issue is frequency dependence. The quality factor $Q(\omega)$ is given by $Q(\omega) = L\omega/R$ with $R$ being the resistance and $\omega$ the angular frequency. The $\omega$-dependence of $L$ in ref. (41) is Debye type with the cut-off of the order of 10kHz. This limits the value of $Q(\omega)$, and a wider range of frequency is needed for the applications. The third one is the nonlinearity with respect to the current density. In ref.(41), the tilt angle $\phi$ is expanded in the current density $j$ as $\phi = Aj + Bj^3 + Cj^5$, and this phenomenological expression well describes the experimental result. However, the microscopic understanding of this nonlinearity is missing. Recently, Ieda and Yamane (41) studied a related problem taking into account the Rashba spin-orbit interaction together with the nonadiabatic $\beta$-term. They found the sign change of $L$ from positive to a negative value as $\beta$ increases.

**Results**
- **Models**

In this paper, we study the microscopic model of spiral magnet composed of one-dimensional electrons coupled with localized spins by exchange interaction. The spiral order occurs at $Q = 2k_F$ with $k_F$ being the Fermi wavenumber of the conduction electrons. This can be regarded as the Ruderman-Kittel-Kasuya-Yosida (RKKY) interaction (25,26) or the Peierls instability (42,43). Assuming that $Q$ is incommensurate with the original lattice, the spiral order breaks two kinds of symmetries, i.e., the translational symmetry and the SU(2) spin rotational symmetry. The resulting symmetry is the combination of these two, i.e., the translation combined with the spin rotation, and hence the number of the Goldstone boson is 3 in the absence of the spin-orbit interaction. Two of which are the fluctuation of the plane of the spin rotation, and the last one is the so-called phason corresponding to $\psi$ in the expression of the director $\boldsymbol{n}$ of the localized spin which acts as the order parameter;

$$\boldsymbol{n} = \boldsymbol{\eta}_1 cos(\boldsymbol{Q} \cdot \boldsymbol{r} + \psi) + \boldsymbol{\eta}_1 sin(\boldsymbol{Q} \cdot \boldsymbol{r} + \psi) + \boldsymbol{\eta}_3 m_z, \qquad (3)$$

where $\boldsymbol{\eta}_i$ is a unit vector in the Cartesian coordinates. Here note that the uniform spin component $m_z$ perpendicular to the spin rotating plane is the generator of $\psi$ corresponding to the "momentum" of the "coordinate" $\psi$. The basic idea of the present paper is that the inductance $L$ of the system is related to the imaginary part of the complex impedance $Z(\omega)$, which is inverse of the conductance $\Sigma(\omega)$. Both the spin transfer torque and the resultant emergent electric field are included in the conductance $\Sigma(\omega)$ due to the collective modes of the spiral spins, i.e., the uniform magnetization and phason. Here some remarks are in order about the difference between the spiral magnet and the conventional collinear spin-density wave (SDW), which share a similar phason collective mode. One is that the system remains metallic even in the perfectly nested case for the spiral state while it is gapped in the collinear SDW. In the latter case, the broken symmetry is the translational symmetry and the spin rotational symmetry. The Goldstone mode corresponding to the former is the phason, while that for the latter is the spin wave. These two are decoupled in the bilinear order, and only the phason contributes to the conductivity. The phason is usually pinned by the impurity, showing the finite pinning frequency of its spectrum. In sharp contrast, the phason in spiral magnet remains gapless even with the disorder as long as the spin rotational symmetry of the Hamiltonian is intact. Both the impurity/commensurability and the spin-orbit are needed to gap the phason spectrum there.

As a microscopic model of electrons coupled to a spiral spin order, we have considered

$$H = \int \frac{dk}{2\pi} \left[ c_k^\dagger \left( \frac{k^2}{2m_e} - E_F - Jm_z(t)\sigma_z \right) c_k - eA(t) c_k^\dagger v_k c_k \right.$$
$$\left. -J \left( m_Q(t) c_{k+\frac{Q}{2}}^\dagger \sigma c_{k-\frac{Q}{2}} + m_Q^*(t) c_{k-\frac{Q}{2}}^\dagger \sigma c_{k+\frac{Q}{2}} \right) \right], \quad (4)$$

where $c_k$ is a Fermion annihilation operator with momentum $k$, $E_F = \frac{Q^2}{8m_e}$ is the Fermi energy, $m_e$ is an effective electron mass, $J$ is an exchange constant, $\sigma$ is the Pauli matrix vector corresponding electron's spin operators, $A(t)$ is an electromagnetic vector potential, $m_Q(t) = \frac{x+y}{2} e^{i\psi(t)} \approx \frac{x+y}{2}[1 - i\psi(t)]$ characterizes the spiral magnetic order whose wave vector is the magnitude of $Q = 2k_F$ and a spin rotation plane is $s_x - s_y$ plane, $\psi(t)$ is a dynamical phase degrees of freedom, namely phason, and $m_z(t)$ is a uniform moment. Note that we have considered one-dimensional free electrons just for simplicity; however, our analysis is general and applicable for higher dimensions and systems with spin-orbit coupling. By introducing the spinor, $\Psi_k = [c_{k+Q/2,\uparrow}, c_{k-Q/2,\downarrow}]^T$, the Hamiltonian is simplified as

$$H = \int \frac{dk}{2\pi} \Psi_k^\dagger \left[ h_0(k) + jA(t) + S_\psi \psi(t) + S_z m_z(t) \right] \Psi_k \quad (5)$$

where $h_0(k) = \frac{k^2}{2m_e}\tau_0 + \frac{Qk}{2m_e}\tau_z - J\tau_x$ is the unperturbed mean field Hamiltonian, $j_k = -e\left(\frac{k}{m_e}\tau_0 + \frac{Q}{2m_e}\tau_z\right)$ is a current operator, $S_\psi = J\tau_y$ and $S_z = -J\tau_z$ are spin operators coupled to phason and uniform magnetization, respectively, and $\tau$ is a vector of the Pauli matrix. The eigenvalues of $h_0(k)$ are given by $\xi_{k,\pm} = \frac{k^2}{2m_e} \pm \sqrt{\left(\frac{Qk}{2m_e}\right)^2 + J^2}$, as shown in FIG. 1. As mentioned, the dispersion retains the metallic Fermi surfaces in addition to the gap of $\Delta = 2J$ at $k = 0$ due to nesting of the Fermi surface. With a basis diagonalizing the mean field Hamiltonian $h_0(k)$, the current and the spin operators are expressed as

$$\tilde{j}_k = -\frac{e}{m} \begin{bmatrix} k - \frac{Q^2 k}{4m_e\zeta} & -\frac{JQ}{2\zeta} \\ -\frac{JQ}{2\zeta} & k + \frac{Q^2 k}{4m_e\zeta} \end{bmatrix}, \tilde{S}_z = \frac{J}{\zeta}\begin{bmatrix} \frac{Qk}{2m_e} & J \\ J & -\frac{Qk}{2m_e} \end{bmatrix}, \tilde{S}_\psi = J\begin{bmatrix} 0 & -i \\ i & 0 \end{bmatrix}, \quad (6)$$

where $\zeta \equiv \sqrt{\left(\frac{Qk}{2m_e}\right)^2 + J^2}$.

As a model describing the magnetic excitation, let us consider the following Lagrangian for the uniform moment and the phason,

$$L_M = -m_z(t)\dot{\psi}(t) - \frac{K_z}{2}m_z^2(t) - \frac{K_\psi}{2}\psi^2(t) - l_a\langle S_z\rangle(t)m_z(t) - l_a\langle S_\psi\rangle(t)\psi(t), \quad (7)$$

where the first term is the Berry phase term representing the canonical conjugate relation between the phason $\psi(t)$ and the uniform magnetization $m_z(t)$, the second and the third terms are mass terms of excitations with a gap size of $K_\psi$ and $K_{m_z}$, and $l_a$ is a lattice constant. Each excitation gap, $K_z$ and $K_\psi$, corresponds to the intrinsic and the extrinsic frequency of the spiral, respectively. Note that the excitation gap of phason $K_\psi$ is originally zero as the phason is a Goldstone mode associated with the spontaneous breaking of translational and spin rotation symmetries. Therefore, the excitation gap of phason becomes finite when magnetic impurities or nonmagnetic impurities with spin-orbit interaction are present. The last two terms describe coupling to itinerant electron's spin density. In addition, the Rayleigh dissipation function, $R_d = \frac{\alpha}{2}\left[\dot{m}_z^2(t) + \dot{\psi}^2(t)\right]$, introduces the dissipation to magnetic excitations where $\alpha$ is the Gilbert damping constant.

- **Emergent inductance and its $\omega$-dependence**

The physical process describing the emergent inductance is described by the Feynman diagram in Fig.2 (44,45). The right bubble corresponds to the spin accumulation induced by the external electric fields, including the spin transfer torque effect; the electric drives the collective modes of the spins. The left bubble corresponds to the emergent electric field, i.e., the collective modes affect the motion of the conduction electrons. The combination of these two processes contributes to the conductivity of the total system, and below, we discuss each of them separately. See Eq.(22) below. Within the linear response theory, current density induced by magnetic excitations are given as

$$\langle j\rangle(\Omega) = \sigma_{m_z}(\Omega)m_z(\Omega) + \sigma_\psi(\Omega)\psi(\Omega), \quad (8)$$

where $\Omega$ is an external frequency, and $\sigma_{m_z}(\Omega)$ and $\sigma_\psi(\Omega)$ are conductivity related to ferromagnetic and phason excitation, respectively. Each conductivity is evaluated as

$$\sigma_{m_z}(\Omega) = -\frac{1}{2\pi\beta}\int dk \sum_{\omega_n}\text{Tr}\left[jG_k(\omega_n+\Omega_m)S_zG_k(\omega_n)\right]\Big|_{i\Omega_m\to\Omega+i0}$$

$$\approx -\frac{i\tau_e\Omega}{2\pi}\sum_{i=\pm}\int dk(\tilde{j}_k)_{ii}(\tilde{S}_z)_{ii}f'(\xi_{k,i}) = -i\Omega C_{m_z}(Q), \quad (9)$$

$$\sigma_\psi(\Omega) = -\frac{1}{2\pi\beta}\int dk \sum_{\omega_n}\text{Tr}\left[jG_k(\omega_n+\Omega_m)S_\psi G_k(\omega_n)\right]\Big|_{i\Omega_m\to\Omega+i0}$$

$$\approx \Omega\sum_{i\neq j}\int\frac{dk}{2\pi}\frac{(\tilde{j}_k)_{ij}(\tilde{s}_\psi)_{ji}-(\tilde{s}_\psi)_{ij}(\tilde{j}_k)_{ji}}{(\xi_{k,i}-\xi_{k,j})^2}f(\xi_{k,i}) = -i\Omega C_\psi(Q) \quad (10)$$

where $G_k(\omega_n) = \left[i\omega_n - h_0(k)\right]^{-1}$ is a Green's function, $\tau_e$ is a scattering lifetime, $f(x)$ is the Fermi-Dirac distribution function, and $C_{m_z}(Q)$ and $C_\psi(Q)$ are coefficients characterizing uniform moment and phason conductivity, respectively. The analytical expressions of the coefficients are given by $C_{m_z}(Q) = \frac{eQ\tau_e}{4\pi\varsigma'\sqrt{m_e}}\sqrt{\varsigma' - \frac{Q^2}{2m_e}}\left(\varsigma' + \frac{Q^2}{2m_e}\right)$ and $C_\psi(Q) = \frac{eQ}{2\pi\sqrt{m_e}}\sqrt{\varsigma' + \frac{Q^2}{2m_e}}^{-1}$ where $\varsigma' = \sqrt{\left(\frac{Q^2}{2m_e}\right)^2 + 4J^2}$. When $\frac{Q^2}{2m_e} \ll 2|J|$ corresponding to long pitch spirals or the adiabatic limit, the coefficients are simplified as $C_{m_z} = \frac{e\tau_e JQ}{\pi\sqrt{8Jm_e}}$ and $C_\psi = \frac{eQ}{\pi\sqrt{8Jm_e}}$, which is linear in $Q$. On the other hand, for the short pitch spiral or the nonadiabatic limit where $\frac{Q^2}{2m_e} \gg 2|J|$ is satisfied, the coefficients are given as $C_{m_z} = \frac{eJ\tau_e}{\pi}, C_\psi = \frac{e}{2\pi}$, which is independent of $Q$. Note, however, that our analysis is basically the random phase approximation, which is justified in the weak coupling limit, i.e., $\frac{Q^2}{2m_e} \gg 2|J|$. Also, since our primary interests are in the nonadiabatic limits, we focus on $Q \to \infty$ limits in the rest of the manuscripts. In the equality between the first and second lines in Eq.(9) and (10), we have performed the analytical continuation, $i\Omega_m \to \Omega + i0$. We have also expanded the response function with respect to the external frequency, $\Omega$, up to the first order; however, the zeroth-order in $\Omega$ vanishes. Because $\tilde{j}_k$ and $\tilde{S}_z$ are symmetric matrices, the conductivity related to ferromagnetic excitation, $\sigma_{m_z}$, is only given by diagonal elements of $(\tilde{j}_k)_{ij}(\tilde{S}_z)_{ji}$; only states near the Fermi surface contribute to $\sigma_{m_z}$. In contrast, the phason spin operator $\tilde{S}_\psi$ only consists of off-diagonal elements, namely, only inter-band transitions contribute to the phason conductivity. This difference is reflected in the lifetime dependence of the conductivities; $\sigma_{m_z}$ is linear to $\tau_e$ whereas $\sigma_\psi$ is independent. Finally, the current density induced by magnetic excitation is summarized as

$$\langle j \rangle(\Omega) = -i\Omega C_{m_z} m_z(\omega) - i\Omega C_\psi \psi(\omega) - i\Omega \frac{eJ\tau_e}{\pi} m_z(\omega) - i\Omega \frac{e}{2\pi}\psi(\omega) \quad (11).$$

Then, let us consider the spin densities induced by applied electric fields,

$$\langle S_\psi \rangle(\Omega) = \chi_\psi(\Omega) A(\Omega), \quad (12)$$
$$\langle S_z \rangle(\Omega) = \chi_{m_z}(\Omega) A(\Omega), \quad (13)$$

where $\chi_{m_z}$ and $\chi_\psi$ are electromagnetic susceptibilities, and $\langle S_\psi \rangle$ and $\langle S_z \rangle$ are spin densities coupled to phason and uniform magnetization, respectively. Similar to the conductivity, the susceptibilities are evaluated as

$$\chi_\psi(\Omega) = -\frac{1}{2\pi\beta}\int dk \sum_{\omega_n} \text{Tr}\left[S_\psi G_k(\omega_n + \Omega_m) j G_k(\omega_n)\right] = -\sigma_\psi(\Omega), \quad (14)$$

$$\chi_{m_z}(\Omega) = -\frac{1}{2\pi\beta}\int dk \sum_{\omega_n} \text{Tr}\left[S_z G_k(\omega_n + \Omega_m) j G_k(\omega_n)\right] = \sigma_{m_z}(\Omega). \quad (15)$$

Similar to the relation between spin-transfer torque and the spin motive force (46), the spin densities induced by applied electric fields are related to the current response driven by magnetic excitations. It is worth mentioning that the phason contribution has an opposite sign with conductivity, whereas the contribution from uniform magnetization has the same sign. The difference attributes to the fact that the phason response comes from an inter-band contribution while the uniform moment excitation comes from the intra-band contribution. In other words, the uniform moment contribution, $C_{m_z}(Q)$, is a transport-like contribution and the phason contribution, $C_\psi(Q)$, is a geometric contribution. Finally, the spin densities induced by an applied electric field is obtained as

$$\langle S_\psi \rangle(\Omega) = i\Omega C_\psi A(\Omega) = -\frac{e}{2\pi}E(\Omega), \quad (16)$$

$$\langle S_z \rangle(\Omega) = -i\Omega C_{m_z} A(\Omega) = \frac{eJ\tau_e}{\pi}E(\Omega). \quad (17)$$

As current and magnetic responses in itinerant electrons have been evaluated, the magnetic dynamics under an applied electric field are considered in the following. Applying the Euler-Lagrange equation, $\frac{\delta L_M}{\delta q} - \frac{d}{dt}\frac{\delta L_M}{\delta \dot{q}} - \frac{\delta R_d}{\delta \dot{q}} = 0$, to the Lagrangian, Eq.(7), and the Rayleigh dissipation function, we obtain the equations of motions for $m_z$ and $\psi$ as

$$\dot{m}_z(t) - K_\psi \psi(t) - \alpha\dot{\psi}(t) - l_a \langle S_\psi \rangle(t) = 0, \quad (18)$$

$$-\dot{\psi}(t) - K_z m_z(t) - \alpha\dot{m}_z(t) - l_a \langle S_z \rangle(t) = 0. \quad (19)$$

A solution of the equations are obtained as

$$\begin{bmatrix} \psi(\omega) \\ m_z(\omega) \end{bmatrix} = l_a D(\omega) \begin{bmatrix} \langle S_\psi \rangle(\omega) \\ \langle S_z \rangle(\omega) \end{bmatrix}, \quad (20)$$

$$D(\omega) = \begin{bmatrix} -K_\psi + i\alpha\omega & -i\omega \\ i\omega & -K_z + i\alpha\omega \end{bmatrix}^{-1}, \quad (21)$$

where we have shifted to a frequency representation and $D(\omega)$ is a Green's function for magnetic excitations.

By equating Eq.(11), (16), (17), and (20), the current density is evaluated as $\langle j \rangle(\omega) = \Sigma(\omega) E(\omega)$ where a complex conductivity is given by

$$\Sigma(\omega) = \boldsymbol{\sigma}_m^T(\omega) D(\omega) \boldsymbol{\chi}_m(\omega)/l_a \quad (22)$$

where $\boldsymbol{\sigma}_m = [\sigma_\psi, \sigma_{m_z}]^T$ and $\boldsymbol{\chi}_m = [\chi_\psi, \chi_{m_z}]^T$. Here, we have converted the one-dimensional current density to the three-dimensional current by $\langle j_{3D} \rangle = \langle j_{1D} \rangle/l_a^2$. The impedance $Z(\omega)$ is determined as an inverse of the complex conductivity, $Z(\omega) = \frac{l_s}{A}[\sigma_{dc} + \Sigma(\omega)]^{-1}$ where $\sigma_{dc} = \frac{e^2 Q\tau}{2\pi m_e l_a^2}$ is the DC conductivity which is irrelevant to the magnetic excitation, $A$ is a cross section of a system, and $l_s$ is a length between electrodes. Then the complex inductance is defined as $L(\omega) = Z(\omega)/(-i\omega)$. In the absence of impurity pinning where the phason excitation is gapless, $K_\psi = 0$, a real part of the inductance, as shown in FIG.3 (A), takes positive value when $\omega \ll \omega_{int}$ where $\omega_{int} = \alpha K_z$ is an intrinsic pinning frequency corresponding to the excitation of uniform moments, while rapidly decreases above $\omega_{int}$. At the same time, the imaginary part of the inductance peaks at $\omega_{int}$, showing the characteristic behavior of the Debye-type relaxation. In contrast, when impurity pinning is present, an additional peak structure at $\omega = \omega_{ext}$ where $\omega_{ext} \sim \sqrt{K_\psi K_z}$ associated with the phason excitation appears on the imaginary part of inductance as shown in FIG.3 (B). The real part of the inductance is negative with the typical parameters when $\omega < \omega_{ext}$, while showing the sign change above $\omega_{ext}$.

In the low frequency limit, the impedance can be expanded as $Z(\omega) \approx \frac{l_s}{\sigma_{dc}+\text{Re}[\Sigma(\omega)]} - i\omega l_s \frac{\text{Im}[\Sigma(\omega)/\omega]}{(\sigma_{dc}+\text{Re}[\Sigma(\omega)])^2}$. Compering with the conventional expression, $Z(\omega) = R - i\omega L$, where $R$ is resistance and $L$ is inductance, one can regard the imaginary part of the complex conductivity as the inductance of spiral;

$$L = l_s \left.\frac{\text{Im}[\Sigma(\omega)/\omega]}{(\sigma_{dc}+\text{Re}[\Sigma(\omega)])^2}\right|_{\omega \to 0}. \quad (23)$$

Note that the equivalence of the imaginary part of the conductivity and the inductance is valid when the real part of the conductivity is larger than its imaginary part, namely $\sigma_{dc} + \text{Re}[\Sigma(\omega)] \gg \text{Im}[\Sigma(\omega)]$. Finally, the complex conductivity $\Sigma(\omega)$ and the inductance with a gapless phason are obtained as

$$\Sigma(\omega) = \frac{C_\psi^2}{l_a \alpha} + i\omega \frac{(C_\psi + \alpha C_{m_z})^2}{l_a \alpha^2 K_z} = \frac{e^2}{4\pi^2 l_a \alpha} + i\omega \frac{e^2}{4\pi^2} \frac{(1+2\alpha\tau_e J)^2}{l_a \alpha^2 K_z}, \quad (24)$$

$$L = \frac{l_s e^2}{4\pi^2 l_a A \left(\sigma_{dc} + \frac{e^2}{4\pi^2 l_a \alpha}\right)^2} \frac{(1+2\alpha\tau_e J)^2}{\alpha^2 K_z}. \quad (25)$$

The expression suggests that both the real and imaginary parts of the complex conductivity are always positive. As a result, the sign of the inductance is also positive when the phason excitation is gapless. On the other hand, in the gapped regime, the conductivity and the inductance mediated by magnetic excitations are obtained as

$$\Sigma(\omega) = \frac{i\omega}{l_a}\left(\frac{C_{m_z}^2}{K_z} - \frac{C_\psi^2}{K_\psi}\right) = \frac{i\omega}{l_a} \frac{e^2}{4\pi^2}\left(\frac{4\tau_e^2 J^2}{K_z} - \frac{1}{K_\psi}\right), \quad (26)$$

$$L = \frac{l_s e^2}{4\pi^2 l_a A \sigma_{dc}^2}\left(\frac{4\tau_e^2 J^2}{K_z} - \frac{1}{K_\psi}\right). \quad (27)$$

The result shows that the imaginary part of the conductivity can take both the positive and negative signs when the magnetic excitation is gapped. Equation (26) consists of two terms; the first term is described by the response function and the characteristic frequency of uniform moment excitations, while those of phason excitation give the second term.

Although $m_z$ and $\psi$ are not independent excitations, the result can be seen as uniform moment excitations, and the phason contributions compete to determine the overall sign of the inductance. Let us refer these two contributions to a $m_z$-contribution and a $\psi$-contribution in the rest of the manuscript. The result in Eq.(27) has some similarity to the result presented in ref. (40), although an approach is different; their analysis is based on spin transfer torques and spin motive force while we have evaluated the complex conductivity based on a microscopic linear response theory.

- **Depinning transition and nonlinear effects**

Now we consider the effect of the depinning of phason on the conductance. It is known that there is a threshold field strength (47) to drive magnetic spirals. The dynamics of the texture is confined around pining centers under the threshold field; the inductance for the pinned magnetic spiral, Eq.(27), is expected. On the other hand, above the threshold field strength, the phason starts freely moving and pinning potential becomes negligible. In this strong field regime, the inductance for gapless phason, Eq.(25), is expected. We have substituted magnetic potentials in Eq.(7) to periodic potentials as $\frac{K_z}{2}m_z^2 \to \frac{K_z}{4}(1 - cos2m_z)$ and $\frac{K_\psi^2}{2} \to K_\psi(1 - cos\psi)$. The substitution introduces the finite potential depth so that the depinning transition can be discussed. We have numerically solved the equation of motion, Eq.(18) and (19), with the nonlinear potentials and perform the Fourier transformation to obtain $\psi(\omega)$ and $m_z(\omega)$, then evaluate current density by Eq.(11). The applied electric field amplitude $E_0$ dependence of the emergent inductance is shown in FIG.4 (A). In the weak field regime, the sign of the inductance is negative, corresponding to pinned magnetic excitations. By increasing the field strength, the emergent inductance decreases quadratically to the field strength; $L = c_0 + c_2 E_0^2$ where $c_0$ and $c_2$ are coefficients. Within the weak field regime, the dynamics of the phason is confined in a potential valley around $\psi = 0$. Further increasing the field strength, the inductance shows a discontinuous jump and changes its sign, corresponding to the depinning transition. Just above the thresholds, the phason dynamics covers two valleys under the oscillating field. Multiple discontinuities in the higher field follow a change in the number of valleys which the phason moved across. Above the thresholds, the inductance shows several sign changes due to nonlinearity in the potentials. The frequency and field dependence of the emergent inductance is shown in FIG.4 (B). With typical material parameters, the threshold field strength is roughly $E_C \sim 0.1$V/m, although the threshold field increases as frequency. Although there are oscillations attributing to nonlinearity in potentials, the emergent inductance is generally positive above $E_C$. These results suggest that the sign of the emergent inductance can be controlled by the external electric field. As suggested in Eq.(27), the sign of the inductance is determined by the competition of two contributions; however, these contributions are material dependent. Utilizing the nonlinearity provides an additional means to manipulate the sign of the inductance.

- **Quality factor : $Q(\omega)$**

Finally, let us discuss the quality factor. The quality factor $Q(\omega)$, which is often used to evaluate the performance of an inductance, is defined by the ratio of the imaginary part and the real part of a complex impedance, $Q(\omega) = -\text{Im}[Z(\omega)]/\text{Re}[Z(\omega)] \approx \omega L/R$. The frequency dependence of the quality factor is presented in FIG.5 (A), showing that a sign change as the inductance and a peak structure at the characteristic frequency of magnetic excitations. When the extrinsic pinning frequency is smaller than the intrinsic pinning, $\omega_{ext} < \omega_{int}$, the quality factor is negative below the extrinsic pinning frequency while changes to positive above it. On the other hand, when the extrinsic pinning frequency exceeds the intrinsic pinning frequency, the quality factor takes a positive value associated with the extrinsic pinning, then changes to negative. The frequency and the extrinsic pinning frequency dependence of the quality factor is shown in FIG.5 (B), showing that the quality factor becomes larger when $\omega_{ext} > \omega_{int}$. This suggests that the quality factor is improved by intentionally introducing impurities to increase the extrinsic pinning frequency. The quality factor is usually the order of $Q \sim 10^{-1}$ with the typical parameters in spiral magnets, whereas current commercial inductors have the order of magnitude larger quality factor, $Q = 10^1 \sim 10^2$. However, as it is inversely proportional to the resistance, a higher quality factor is expected for metals with higher mobility.

**Discussion**

Experimentally, negative inductance has been observed in a short-period helimagnet $Gd_3Ru_4Al_{12}$. Our result given in Eq.(27) suggests that, in $Gd_3Ru_4Al_{12}$, magnetic excitation is gapped and the phason contribution is considered to be dominant comparing to contributions from uniform moments. The Debye-type relaxation behavior shown in FIG. 3 is also observed in $Gd_3Ru_4Al_{12}$. Regarding the nonlinearity, our result given in FIG.4 shows quadratic dependence to an applied field, $L = c_0 + c_2 E_0^2$; the same behavior was experimentally reported in ref. (41). Recently, another short-pitch metallic helimagnet $YMn_6Sn_6$ is reported to show considerably large emergent inductance above the room temperature (48). In this compound, the inductance changes its sign from negative to positive as increasing temperature towards the phase transition temperature to the forced ferromagnetic states. This behavior can be explained by the softening of the phason modes; in the low temperature regime, the phason is pinned by impurities while, in the high temperature near the transition, the phason is thermally excited and depinned. The sign change of the inductance from negative to positive by increasing applied current density is also reported. Our result shown in FIG. 4 suggests that the sign change is attributed to the depinning transition of the phason.

Let us discuss the $Q$-dependence of the inductance. In the previous studies based on the adiabatic approach: the spin transfer torque and the emergent electromagnetic field (39,40), the inductance is predicted to be proportional to $Q$. Contrary, our results shown in Eq.(25) and (27) are independent of $Q$ when $\frac{Q^2}{2m_e} \gg 2J$. This discrepancy arises from differences in considered limits; the previous studies focus on the long pitch spiral or the adiabatic limit, whereas our focus is on the short pitch spiral or the nonadiabatic limit. Thus, our study and the previous studies are complementary. These results suggest that there is a crossover around $\frac{Q^2}{2m_e} \sim 2J$ between the adiabatic and nonadiabatic transport, and the inductance saturates and becomes independent of the pitch of the spiral as $Q$ increases.

Lastly, we discuss the stability of the system. The negative $L$ does not mean the negative energy for the magnetic field in the present case. Instead, it is related to the electric field with momentum $q = 0$. The stability of the system is ensured by the analyticity of conductance or conductivity in the upper half of the complex frequency plane, i.e., the causality condition is satisfied. For example, in classical electromagnetism, a resistor-inductor circuit (RL circuit) with negative inductance is known to be unstable as its transient solution is given by $I(t) \propto e^{-Rt/L}$ where $I(t)$ is a current. The conductance in this system is given by $G(\omega) = 1/(R - i\omega L)$, whose pole is located at $\omega_p = -iR/L$. When both $R$ and $L$ are positive, the conductance is analytical in the upper half of the complex plane ensuring stability, whereas, when the $L < 0$, the conductance is no longer analytical and the system becomes unstable. To apply the same argument spiral magnets, the pole of the complex conductivity, Eq.(22), are found at

$$\omega_p = \begin{cases} \pm\frac{\sqrt{4K_z K_\psi - \alpha^2(K_z - K_\psi)^2}}{2(1+\alpha^2)} - i\frac{\alpha(K_z + K_\psi)}{2(1+\alpha^2)} & \left(\frac{\alpha^2}{4} < \frac{K_z K_\psi}{(K_z - K_\psi)^2}\right) \\ -i\frac{\sqrt{\alpha^2(K_z - K_\psi)^2 - 4K_z K_\psi} \pm \alpha(K_z + K_\psi)}{2(1+\alpha^2)} & \left(\frac{\alpha^2}{4} > \frac{K_z K_\psi}{(K_z - K_\psi)^2}\right), \end{cases} \quad (28)$$

where all the poles are located in the lower-half complex plane ensuring analyticity of the conductivity in the upper-half of the complex plane. Namely, the stability of the system is retained even with a negative inductance in spiral magnets.

In conclusion, we have microscopically derived the analytical expressions of an emergent inductance in spiral magnets based on the linear response theory of conductivity and identified the role of the phason collective modes in the emergent induction. We revealed that the sign of inductance is positive when the phason excitation is gapless in the absence of impurity pining, while it can be both negative and positive in the presence of pinning. For the pinned case, the sign of inductance is determined by a competition between contributions from phason and uniform magnetization excitations; phason excitation contributes to negative inductance and uniform moment excitation contributes to positive inductance. We further investigate the nonlinearity in emergent inductance and found that depinning transition of the magnetic spiral can cause the sign change of inductance. This nonlinearity and depinning processes provide a way to control the sign of inductance by external electric fields. Finally, we have evaluated the quality factor and its dependence on extrinsic pinning. As a result, we found that the quality factor becomes larger when the extrinsic pinning potential exceeds the intrinsic pinning potential. We believe that our results would provide the microscopic understanding and essential knowledge for devise applications of the emergent inductance.

**Acknowledgments**
The authors are grateful to J. Ieda, Y. Yamane, K. Yamamoto, S. Maekawa, Y. Tokura, M. Hirschberger, A. Kitaori, and Y. Yokouchi for insightful discussions.

**Funding**
D.K. was supported by the RIKEN Special Postdoctoral Researcher Program. N.N. was supported by JST CREST Grant Number JPMJCR1874 and JPMJCR16F1, Japan, and JSPS KAKENHI Grant numbers 18H03676.

**Author contributions**
D.K. constructed the theory, conducted the calculations, analyzed the data, and wrote the paper. N.N. conceived the study, constructed the theory, and wrote the paper.

**Data and materials availability**
All data needed to evaluate the conclusions in the paper are present in the paper. Additional data related to this paper may be requested from the author.

**Competing interests**
The authors declare that they have no competing interests.

**Figures and Tables**

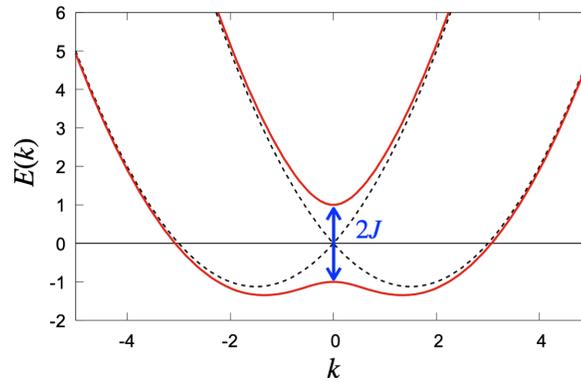

**Fig. 1. Band Dispersion of one-dimensional electrons with spiral SDW order.** Parameters are taken as $m_e = 1, J = 1$ and $Q = 3$.

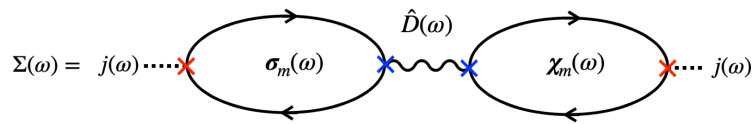

**Fig. 2. Feynman diagram for magnetic excitation mediated conductivity.** The electron bubbles corresponds to a current-spin correlation functions. The solid and wave lines are propagators of itinerant electrons and magnetic excitations, respectively.

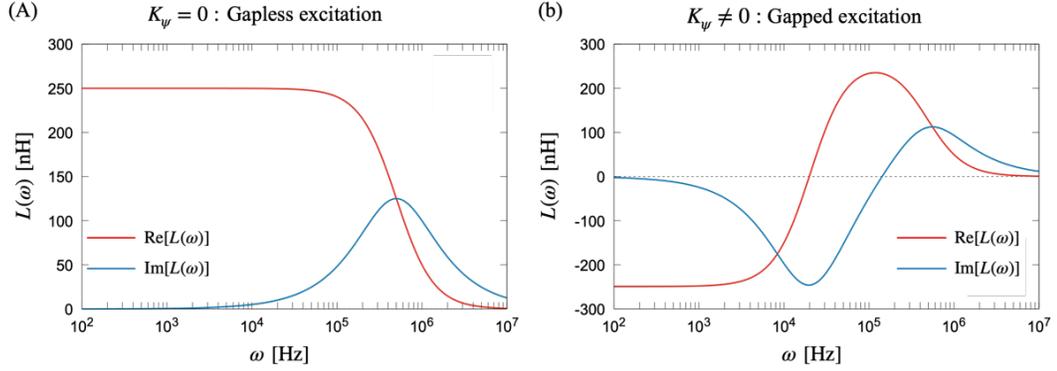

**Fig. 3. Frequency dependence of the emergent inductance $L(\omega)$.** (A) The inductance with the gapless magnetic excitation, $K_\psi = 0$. (B) The inductance with the gapped magnetic excitation, $K_\psi = 10^4$Hz. Parameters are taken as $K_z = 10^6$Hz, $J = 200$meV, $\tau_e = 10$fs, $A = 10\mu m^2$, $l_a = 4$Å, $l_s = 10\mu m$, $\alpha = 0.5$, and $Q = 2.24$nm$^{-1}$. The DC Drude conductivity with these parameters is evaluated as $\sigma_{dc} = 6.3 \times 10^5 \Omega^{-1}m^{-1}$.

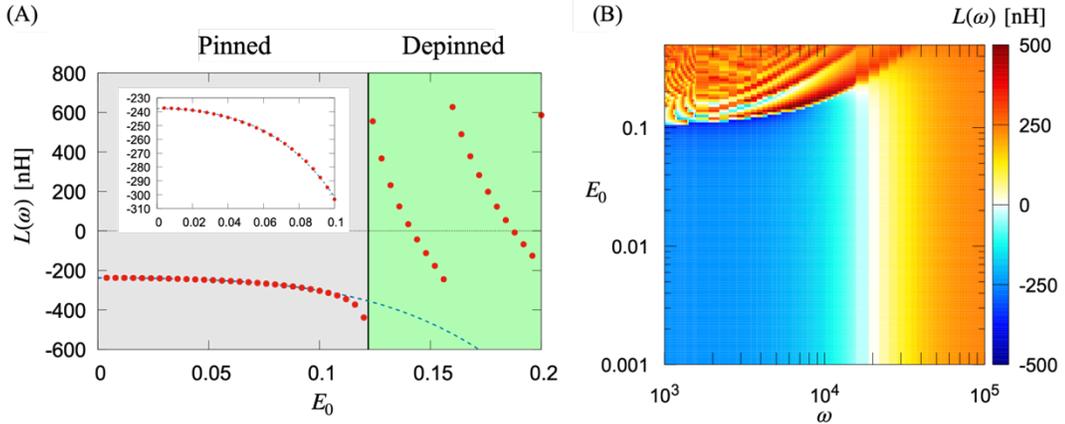

**Fig. 4. Nonlinear effects on the inductance.** (A) The applied electric field dependence of the emergent inductance $L(\omega)$ at $\omega = 3$kHz. The inset shows the magnified behavior at the low field region. (B) Frequency and field strength dependence of the emergent inductance. Parameters are taken as $K_z = 10^6$Hz, $K_\psi = 10^4$ Hz, $J = 200$meV, $\tau_e = 10$fs, $A = 10\mu m^2$, $l_a = 4$Å, $l_s = 10\mu m$, $\alpha = 0.5$, and $Q = 2.24$nm$^{-1}$.

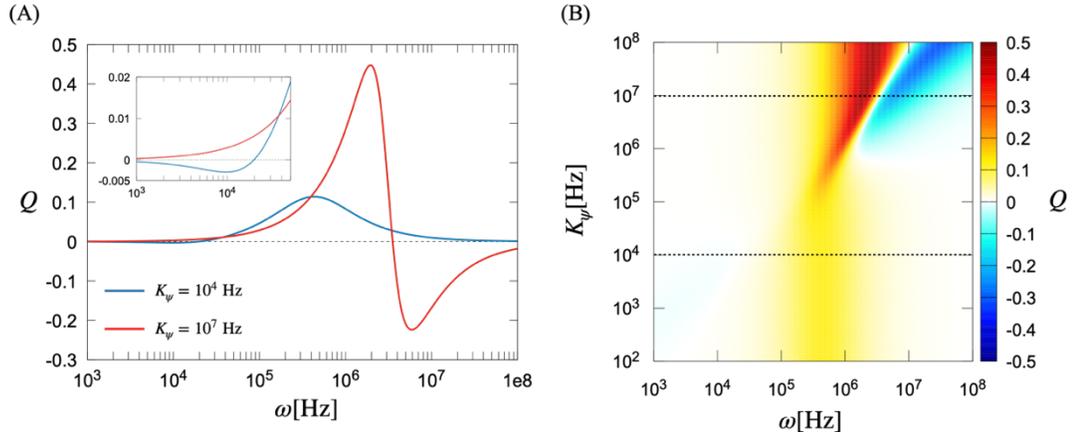

**Fig. 5. Quality factor.** (A) Frequency dependence of the quality factor with the phason gap $K_\psi = 10^4$Hz and $10^7$Hz. The inset shows a magnified behavior at the low frequency regime. (B) Frequency and phason gap $K_\psi$ dependence of the quality factor. Parameters are taken as $K_z = 10^6$ Hz, $J = 200$meV, $\tau_e = 10fs$, $A = 10\mu m^2$, $l_a = 4$Å, $l_s = 10\mu m$, $\alpha = 0.5$, and $Q = 2.24$ nm$^{-1}$. The DC Drude conductivity with these parameters is evaluated as $\sigma_{dc} = 6.3 \times 10^5 \Omega^{-1} m^{-1}$.